\documentstyle[12pt]{article}
\newcommand{\nc}{\newcommand}

\nc{\dbar}{\bar{\partial}}

\catcode`\@=11

\@addtoreset{equation}{section}
\def\theequation{\thesection\arabic{equation}}

\def\@normalsize{\@setsize\normalsize{15pt}\xiipt\@xiipt
\abovedisplayskip 14pt plus3pt minus3pt%
\belowdisplayskip \abovedisplayskip
\abovedisplayshortskip  \z@ plus3pt%
\belowdisplayshortskip  7pt plus3.5pt minus0pt}
\def\small{\@setsize\small{13.6pt}\xipt\@xipt
\abovedisplayskip 13pt plus3pt minus3pt%
\belowdisplayskip \abovedisplayskip
\abovedisplayshortskip  \z@ plus3pt%
\belowdisplayshortskip  7pt plus3.5pt minus0pt
\def\@listi{\parsep 4.5pt plus 2pt minus 1pt
            \itemsep \parsep
            \topsep 9pt plus 3pt minus 3pt}}

\def\underline#1{\relax\ifmmode\@@underline#1\else
        $\@@underline{\hbox{#1}}$\relax\fi}
\@twosidetrue
\relax

\catcode`@=12

\catcode`\@=11

\def\section{\@startsection{section}{1}{\z@}{3.5ex plus 1ex minus
   .2ex}{2.3ex plus .2ex}{\large\bf}}

\def\ps@headings{\def\@oddfoot{}\def\@evenfoot{}
\def\@oddhead{\hbox{}\hfill
        \makebox[.5\textwidth]{\raggedright\ignorespaces --\thepage{}--
        \hfill }}
\def\@evenhead{\@oddhead}
\def\subsectionmark##1{\markboth{##1}{}}
}

\ps@headings

\catcode`\@=12

\relax

\def\figcap{\section*{Figure Captions\markboth
        {FIGURECAPTIONS}{FIGURECAPTIONS}}\list
        {Fig. \arabic{enumi}:\hfill}{\settowidth\labelwidth{Fig. 999:}
        \leftmargin\labelwidth
        \advance\leftmargin\labelsep\usecounter{enumi}}}
 \relax
\def\tablecap{\section*{Table Captions\markboth
        {TABLECAPTIONS}{TABLECAPTIONS}}\list
        {Table \arabic{enumi}:\hfill}{\settowidth\labelwidth{Table 999:}
        \leftmargin\labelwidth
        \advance\leftmargin\labelsep\usecounter{enumi}}}
 \relax
\def\reflist{\section*{References\markboth
        {REFLIST}{REFLIST}}\list
        {[\arabic{enumi}]\hfill}{\settowidth\labelwidth{[999]}
        \leftmargin\labelwidth
        \advance\leftmargin\labelsep\usecounter{enumi}}}
 \relax

\catcode`\@=11

\def\ps@headings{\def\@oddfoot{}\def\@evenfoot{}
\def\@oddhead{\hbox{}\hfill
        \makebox[.5\textwidth]{\raggedright\ignorespaces --\thepage{}--
        \hfill }}
\def\@evenhead{\@oddhead}
\def\subsectionmark##1{\markboth{##1}{}}
}

\ps@headings

\relax

\def\firstpage#1#2#3#4#5#6{
\begin{document}

\begin{titlepage}
\nopagebreak
\title{\begin{flushright}
      \vspace*{-1.3in}
        {\normalsize  CK-TH-98-003 }\\[-5mm]
{\normalsize December 1998}\\[.5cm]
\end{flushright}
\vspace{10mm}
{\large \bf #3}}
\vspace{1cm}
\author{\large #4 \\ #5}
\maketitle
\vskip -1mm
\nopagebreak
\begin{abstract}
{\noindent #6}
\end{abstract}
\vspace{1cm}
\begin{flushleft}
\rule{16.1cm}{0.2mm}\\[-3mm]
$^{1}${\small E--mail: c.kokorelis@sussex.ac.uk}
\end{flushleft}
\thispagestyle{empty}
\end{titlepage}}
\newcommand{\dal}{\raisebox{0.085cm}
{\fbox{\rule{0cm}{0.07cm}\,}}}
\newcommand{\bb}{\begin{eqnarray}}
\newcommand{\ee}{\end{eqnarray}}
\newcommand{\p}{\partial}
\newcommand{\bp}{{\bar \p}}
\newcommand{\bR}{{\bf R}}
\newcommand{\bC}{{\bf C}}
\newcommand{\bZ}{{\bf Z}}
\newcommand{\bS}{{\bar S}}
\newcommand{\bT}{{\bar T}}
\newcommand{\bU}{{\bar U}}
\newcommand{\bA}{{\bar A}}
\newcommand{\bh}{{\bar h}}
\newcommand{\bu}{{\bf{u}}}
\newcommand{\bv}{{\bf{v}}}
\newcommand{\D}{{\cal D}}
\newcommand{\s}{\sigma}
\newcommand{\Sg}{\Sigma}
\newcommand{\ket}[1]{|#1 \rangle}
\newcommand{\bra}[1]{\langle #1|}
\newcommand{\non}{\nonumber}
\newcommand{\ph}{\varphi}
\newcommand{\la}{\lambda}
\newcommand{\ga}{\gamma}
\newcommand{\ka}{\kappa}
\newcommand{\m}{\mu}
\newcommand{\n}{\nu}
\newcommand{\th}{\vartheta}
\newcommand{\Lie}[1]{{\cal L}_{#1}}
\newcommand{\eps}{\epsilon}
\newcommand{\bz}{\bar{z}}
\newcommand{\bX}{\bar{X}}
\newcommand{\om}{\omega}
\newcommand{\Om}{\Omega}
\newcommand{\we}{\wedge}
\newcommand{\La}{\Lambda}
\newcommand{\bOm}{{\bar \Omega}}
\newcommand{\CA}{{\cal A}}
\newcommand{\CF}{{\cal F}}
\newcommand{\CbF}{\bar{\CF}}
\newcommand{\CAM}{\CA^{(M)}}
\newcommand{\CAS}{\CA^{(\Sg)}}
\newcommand{\CFS}{\CF^{(\Sg)}}
\newcommand{\I}{{\cal I}}
\newcommand{\al}{\alpha}
\newcommand{\be}{\beta}
\newcommand{\cm}{Commun.\ Math.\ Phys.~}
\newcommand{\pr}{Phys.\ Rev.\ D~}
\newcommand{\pl}{Phys.\ Lett.\ B~}
\newcommand{\ibar}{\bar{\imath}}
\newcommand{\jbar}{\bar{\jmath}}
\newcommand{\np}{Nucl.\ Phys.\ B~}
\newcommand{\e}{{\rm e}}
\newcommand{\gsi}{\,\raisebox{-0.13cm}{$\stackrel{\textstyle
>}{\textstyle\sim}$}\,}
\newcommand{\lsi}{\,\raisebox{-0.13cm}{$\stackrel{\textstyle
<}{\textstyle\sim}$}\,}
\date{}
\firstpage{95/XX}{3122}
{
The Master Equation for the Prepotential} 
{ Christos Kokorelis$^{1}$}
{\normalsize
Centre of Theoretical Physics\\
\normalsize Sussex University\\
\normalsize Falmer, Brighton, BN1 9QH, United Kingdom}
{ 
The perturbative prepotential and the K\"ahler metric of the vector
multiplets of the N=2 effective low-energy heterotic strings is
calculated directly in $N=1$ six-dimensional toroidal
compactifications of the heterotic string vacua.
This method provides the solution for the one loop correction to the
N=2 vector multiplet prepotential for compactifications of
the heterotic string
for any rank three and four models,
as well for compactifications on $K_3 \times T^2$.
In addition, we complete previous calculations, derived from string
amplitudes, by
deriving the differential equation for the third
derivative of the prepotential with respect of the usual
complex structure U moduli of the $T^2$ torus. 
Moreover, we calculate the one loop prepotential, using its modular 
properties, for N=2
compactifications of the heterotic string
exhibiting modular groups similar with those
appearing in N=2 sectors of N=1 orbifolds based on non-decomposable 
torus lattices and on N=2 supersymmetric Yang-Mills. }

\newpage

\section{Introduction}





One of the most important aspects of string dualities involves
comparisons of the effective actions between 
$N=2$ compactifications of the ten dimensional heterotic string to
lower dimensions and type II superstrings.
A key future for testing these dualities 
is the use of prepotential, which describes the $N=2$ effective 
low energy theory of vector multiplets in a general supergravity theory.
Guidance in working, with the vector multiplet heterotic prepotential at the
string theory level, comes from similar results 
from $N=2$ supersymmetric Yang-Mills \cite{seiwi1}.
At the level of $N=2$ supersymmetric $SU(r+1)$ Yang-Mills the
quantum moduli space was associated \cite{club} with a particular genus r
Riemann surface parametrized by r complex moduli and 2r
periods
$(\alpha_{D_{i}},\alpha)$.

In this work we are interested, among the web of dualities, only in the 
proposal of \cite{kava,fhsv} which
provided evidence for the exact nonperturbative equivalence
of dual pairs, on the one side of the 
heterotic string compactified on $K_3 \times T_2$ and on the other side
of the IIA compactified on a Calabi-Yau threefold.
The proposal identifies the moduli spaces of heterotic string and its 
dual IIA as ${\cal M}_V^{heterotic}= 
{\cal M}_V^{IIA}$ and ${\cal M}_H^{heterotic}={\cal M}_H^{IIA}$, where
the subscripts refer to vector multiplets and hypermultiplets 
respectively. 
In other words, for models which are dual pairs,
the exact prepotential for the vector multiplets, including perturbative and
non-perturbative corrections is calculated from the IIA side, where the
tree level result is exact.
On the contrary, if we want to calculate the exact hypermultiplet
superpotential
this is calculated from the heterotic string on $K_3 \times T^2$.
This is exactly Strominger's proposal that the absence of
neutral perturbative couplings between vector multiplets and
hypermultiplets survives nonperturbative string effects.
In this sence the complete prepotentials for the vector
multiplets for the two "different" theories match, including perturbative
and non-perturbative corrections, and ${\cal F}^{het}= {\cal F}^{IIA}$.

Compactification
of the heterotic string on $K_3 \times T_2$ and of type IIA
on a Calabi-Yau threefold produce models
with N=2 supersymmetry in four dimensions.
These models have been tested to be dual to each other,
at the level of effective theories, and for low numbers of
$h^{(1,1)}$ vector multiplet moduli \cite{kava}.
By using mirror symmetry, 
 we can map the vector multiplet sector of IIA to its mirror
type IIB $h^{(2,1)}$ complex structure moduli space\footnote{
Application of the mirror symmetry on the type IIB
interchanges three-cycles with two cycles and three-branes with 
two branes.}.
The procedure then concentrates in the comparison of the 
complex structure moduli \cite{liya} effective theories, of type IIB,
with 
$N=2$ heterotic compactifications of rank three or four 
models \cite{kava,anpa}.
In this sence, in type IIB the result for the effective gauge coupling 
is exact 
to all orders of perturbation theory, since the dilaton in type IIB 
belongs to the hypermultiplet sector and there are no couplings allowed
between \cite{DWLVP} vector multiplets and hypermultiplets.

The paper is organized as follows.
In section two we calculate the one loop correction to the  
perturbative prepotential of the vector multiplets
for the heterotic string compactified on a six dimensional orbifold.
It comes from the general solution from the one loop K\"ahler 
metric \cite{afgnt,anto3}.
Furthermore, we {\em establish a general procedure} for calculating
one loop corrections to the one loop prepotential, not only
for $N=1$ six dimensional heterotic strings toroidally compactified on 
four dimensions,
which has important implications for any compactification of the 
heterotic string in four dimensions having(or not) a type II dual. 
This procedure is an 
alternative way to the calculation of the prepotential which was 
performed indirectly in \cite{mouha} via the effective gauge 
couplings. 
We will calculate directly the integral representation
of $N=2$                                                         
 vector multiplet prepotential of toroidal
compactifications of the 
 heterotic string\footnote{However, due to factorization
properties of the $T^2$ subspace of the heterotic Narain lattice
and instanton embedding independence into the gauge bundle over $K_3$,
the same result can be applied to any heterotic string 
compactification on $K_3 \times T^2$ for any rank four
models.}, continuing the 
work of \cite{anto3,wkll,afgnt}.
The one loop K\"{a}hler metric in the moduli space of 
vector multiplets of toroidal compactifications of $N=1$ six dimensional
orbifold
compactifications \cite{walt,erle} of the heterotic string follows directly
from this result.
Automatically this calculates
one-loop
corrections to the K\"ahler metric for the moduli of
the usual vector multiplet T, U moduli fields of the $T^2$ torus
appearing in $N=2$ $(4,4)$ compactifications of the heterotic
string. The calculation
on the quantum moduli space takes into consideration points of
enhanced gauge symmetry.

In addition our procedure is complementary to \cite{afgnt} since we 
calculate, 
contrary to \cite{afgnt} where the third derivative of the prepotential 
with respect to the T moduli was calculated, the third derivative of the 
prepotential with respect to the complex structure U moduli. 
The logarithmic singularity, consequence of the gauge symmetry enhancement 
at a specific point in the moduli space do appear in the prepotential
and the K\"ahler metric.

In section three we  
describe our results for the calculation 
of
the heterotic prepotential in the
case of toroidal compactifications
of $N{=}1$ six dimensional
orbifold
compactification of the heterotic string, where the 
underlying
torus lattice does not "decompose" as $T^2 \oplus T^4$, the $T^4$ 
could be an orbifold limit of $K_3$.
The moduli of the unrotated complex plane, e.g $T^2$ with 
a shift,  
 has a modular symmetry
group that is a subgroup of $SL(2,Z)$. In particular, we consider
this modular symmetry, e.g $\Gamma^0(3)$,
to be one of those appearing in $N=2$ sectors of non-decomposable 
$N=1$ orbifold compactifications \cite{erkl,deko}.
The calculation of the prepotential is based on the use of its modular 
properties under the action of $\Gamma^0(3)_T \times \Gamma^0(3)_U $
modular
group.

In section four we describe our results for
the $N=2$ prepotential of any rank three compactifications of
the heterotic string.
 A general result \cite{fava1} concerning the geometry, local issues,
behind the existence of heterotic duals, is that 
existence of the weakly coupled heterotic string compactifications on the 
$K_3 \times T^2$ is allowed only when
its type IIA dual is
compactified on a Calabi-Yau manifold which can be written in turn
as
a fibre bundle with base $P^1$ and generic fiber the 
$K_3$ surface \cite{klm1}. At the weak heterotic limit the base $P^1$
goes to infinite size.
We apply the {\em general procedure} for calculating 
the one loop 
heterotic prepotential in a certain rank three heterotic 
model \cite{osa} 
which has a 
type II dual compactified on a Calabi-Yau at a specific
limit.
The Calabi-Yau models incorporate the
 $K_3$ fiber
 structure \cite{klm1}
of the type II dual realization.


\section{ One loop correction to the prepotential from string
amplitudes/ K\"{a}hler metric}

{\em Preliminaries - Rank four models}

The one-loop
K\"{a}hler metric for orbifold compactifications of the heterotic 
string,
where the internal six torus decomposes into $T^2 \oplus T^4$, was 
calculated in \cite{anto3}. In this section, we will use the general 
form of the 
solution for the one loop K\"{a}hler metric appearing 
in \cite{anto3,afgnt}
to calculate the 
one loop correction to the prepotential of $N{=}2$ orbifold 
compactifications of the heterotic string. 
While the one loop prepotential has been calculated with the use 
of string amplitudes in \cite{afgnt}) via its third derivative with
respect to the T moduli, here we will provide an alternative 
way of calculating the one-loop correction \cite{wkll,afgnt}
to the prepotential
of the vector multiplets of the $N{=}2$ orbifold compactifications
of the heterotic string.
For the calculation of the one-loop contribution to the K\"{a}hler metric
we use the linear multiplet formulation \cite{cefergi1,kuue,anto3} and 
not the chiral formulation.
Note that both formulations are equivalent, since the linear multiplet
can always be transformed in to a chiral multiplet by a 
supersymmetric duality transformation.

Lets us suppose that the internal six dimensional lattice 
decomposes into $T^2 \oplus T^4$.
 The $T^4$ part of the lattice, may represent
an orbifold limit\cite{hemo,walt} of $K_3$ while the $T^2$ part,
which contains the usual T, U moduli may contain\footnote{As it happens
in $N=2$ $(4,0)$ models \cite{afgnt,walt}.}
 a lattice shift,
necessary for modular invariance.
In this subspace of the Narain moduli space, we want to calculate the 
moduli dependence
of the one loop correction to the prepotential. 
Denote the untwisted moduli from a $N=2$ sector 
by P, where P can be the T or U moduli
parametrizing \cite{dis} the two dimensional unrotated plane.
Then the one loop contribution \cite{anto3} to the 
 K\"{a}hler metric is given by 
\begin{eqnarray}
G^{(1)}_{P\;{\bar P}}=-\frac{1}{(P - {\bar P})^2} {\cal I},\;\;\;
{\cal I}= \int_{\cal F} \frac{d^2\tau}{{\tau}^2_2} \partial_{\bar \tau}
(\tau_2 Z) {\bar F}({\bar \tau}).
\label{metrik1}
\end{eqnarray}
Here, the integral is over the fundamental domain,
and the factor $-\frac{1}{(P - {\bar P})^2}$ is the tree level moduli 
metric $G^{(0)}_{P\;{\bar P}}$. Z is the partition
function of the fixed torus
\begin{equation}
Z = \sum_{(P_L,P_R) \in \Gamma_{2,2}}
q^{P_L^2 /2} {\bar q}^{P^2_R /2} ,\;q \equiv e^{2 \pi \tau} \;
\tau = \tau_1 + i\tau_2 ,
\label{partik1}                                                                
\end{equation}
and $P_L, P_R$
are the left and right moving momenta associated with this plane.
$F(\tau)$ is a moduli independent meromorphic form\footnote{A 
function f is meromorphic at a point A if 
the function h, $ h(z) \stackrel{def}{=} (z- A) f(z)$ is 
holomorphic (differentiable) at 
the point A. In general, this means that the function h is 
allowed to have poles.}   
of weight $-2$ with a single pole at infinity due to the tachyon
at the bosonic sector.
The function F was fixed in \cite{afgnt} to be 
\begin{equation}
F(\tau)= -(1/\pi)\frac{j(\tau) [j(\tau) -j(i)]}{j_{\tau}(\tau)},\;\;\;
\;\;\;
j_{\tau }\stackrel{def}{=}\frac{\partial j(\tau)}{\partial \tau},
\label{partik2}
\end{equation} 
where j the modular function for the group $SL(2,Z)$.   
The function $F(\tau)$ is actually the index in the Ramond sector in the
in the remaining superconformal blocks. With the use of the relations
between $E_4$, $E_6$, $\triangle$ and j, appearing in our appendix,
 we can easily see that F becomes
\begin{equation}
F(\tau)= -2  i \frac{1}{(2 \pi)^2} \frac{E_4(\tau) E_6(\tau)}
{\triangle(\tau)}.
\label{partiku2341}
\end{equation}
In the above form the index will be used to perform a test on 
different forms of the prepotential as it appears in 
\cite{afgnt} and \cite{mouha}.

{\em  Prepotential of vector multiplets/K\"ahler metric }
 
The convention for the complex dilaton is $<S>= \frac{\theta}{\pi} + i
\frac{8 \pi}{g_s^2}$, where $g_s$ is the four dimensional string
coupling and the $\theta$ angle.
For the calculation of the prepotential of the vector multiplets
we will will follow the approach of \cite{afgnt}.
The space of scalar fields belonging to the
vector multiplet part of the effective low energy theory of 
$N=2$ heterotic string compactifications
on the $K_3 \times T^2$ in four dimensions 
is described by special geometry \cite{sukou,DWLVP}.
Part of the spectrum in four dimensions contains the dilaton and the 
T, U moduli of the $T^2$ torus all belonging to vector multiplets.
The general form of the K\"ahler potential reads
\begin{equation}
K= -\ln (iY),\;Y= 2(F- {\bar F}) - (T -{\bar T})((F_T + 
{\bar F}_{\bar T}) 
-(U -{\bar U})((F_U + {\bar F}_{\bar u})
-(S - {\bar S})(F_S + {\bar F}_{\bar S}),
\label{partiko224}
\end{equation}
\begin{equation}
F = STU + f(T,U)+ f^{non-pert},
\label{partik3}
\end{equation}
with f, the one loop correction to the prepotential F,S the dilaton, and
$f^{NP}$, the non-perturbative contribution
to F. The latter will be neglected in the following
as
we consider the calculation of the prepotential at the semiclassical 
limit $S \rightarrow \infty$.
In the linear multiplet the K\"{a}hler potential may be used 
in the following form
as
\begin{equation}
K=-\ln\{i(S - {\bar S}) -2iG^{(1)} \} + G^{(o)},
\label{partik31}
\end{equation}
where
the quantity $G^{(1)}$ is identified with the Green-Schwarz term 
at the semiclassical weak coupling limit $S \rightarrow \infty$
and $G^{(o)}$ the dilaton independent part.
Expanding ({\ref{partik31}) 
\begin{eqnarray}
K^{(1)}_{P{\bar P}}=\frac{2i}{(S-{\bar S})}G^{(1)}_{P{\bar P}},\;\;
\;\;G^{(1)}_{T\;{\bar T}}=\frac{i}{2(T-{\bar T})^2}
\left(\partial_T -\frac{2}{T-{\bar T}}\right)\left(\partial_U - 
\frac{2}{U - {\bar U}}\right)f + c.c
\label{partik4}
\end{eqnarray}
and
using the equations for the momenta
\begin{equation}
{p_L}=\frac{1}{\sqrt{2ImT ImU}}(m_1+ m_2{\bar U} + n_1 {\bar T}+n_2 
{\bar U}
{\bar T}),\;p_R =\frac{1}{\sqrt{2ImT ImU}}(m_1+ m_2{\bar U}+ n_1 T+
n_2 T{\bar U}),
\label{partik5}
\end{equation}
we can prove that $ \cal I$ has the following 
general solution away of the enhanced symmetry
points\cite{afgnt}
\begin{equation}
{\cal I}=\frac{1}{2i} (\partial_T - \frac{2}{T-{\bar T}})
(\partial_U  - \frac{2}{U - {\bar U}} )f(T,U) + c.c.
\label{partik7}
\end{equation}
The function
f represents the one-loop correction to the prepotential of the 
vector multiplets T, U and determines via eqn.(\ref{partik4}) the one
loop correction to the K\"{a}hler metric for the T, U moduli.
While $f_{TTT}$ was calculated for 
$N=2$ compactifications of the heterotic string in $D=4$
through its modular properties \cite{wkll} and from string 
amplitudes \cite{afgnt}, f could only be calculated 
indirectly \cite{mouha}.
Note that up to now the only way of calculating f for heterotic string
compactifications was indirectly\cite{mouha}, through the
one loop corrections to the Wilsonian gauge couplings \cite{sv} 
coming from the following equation
\begin{equation}
\partial_U \partial_{\bar U} \triangle = bK_{U{\bar U}}^{(o)} +
4 \pi^2 K_{U {\bar U}}^{(1)}.
\label{partik777}
\end{equation}
Here, $\triangle$ is given by \cite{anpatay,dkl2}
is the loop threshold correction to the gauge coupling constants.
In addition, $K_{U{\bar U}}^{(o)}$ is the tree level K\"ahler 
metric \cite{anto3,mouha}
$-1/(U-{\bar U})^2$ and $K_{U {\bar U}}^{(1)}$ is the one 
loop K\"ahler metric 
given by 
\begin{eqnarray}
K_{U {\bar U}}^{(1)}= -\frac{1}{8 \pi^2} \int_{\cal F} \frac{d^2 \tau}
{\tau_2}\partial_U \partial_{\bar U} \left(
  \sum_{
 BPS\;hypermultiplets} - \sum_{ 
 BPS\;vector\;multiplets} \right) e^{i \pi \tau M_L^2} e^{-i \pi {\bar \tau}
  M_R^2},
 \label{kirios1}
 \end{eqnarray}
where $M_L$, $M_R$ are the masses for the left and right movers.

For
$N{=}2$ heterotic strings compactified on decomposable orbifolds
 $f_{TTT}$ is given by \cite{afgnt} 
\begin{equation} 
f_{TTT}= -{2i\over\pi}\frac{j_T(T)}{j(T)-j(U)}
\left\{\frac{j(U)}{j(T)}\right\}
\left\{\frac{j_T(T)}{j_U(U)}\right\}
\left\{\frac{j(U)-j(i)}{j(T)-j(i)}\right\}.
\label{partik10}
\end{equation}
In \cite{wkll,afgnt} $f_{TTT}$ was determined by the property
of behaving as a meromorphic modular form of weight 4 under 
T-duality. 
In addition, $f_{TTT}$ had to vanish at the order 2 fixed point
$U{=}i$ and the order 3 fixed point $U{=}\rho$ of the 
modular group $SL(2,Z)$. Moreover, it had to transform with modular 
weight $-2$ under $SL(2,Z)_U$ transformations and
exhibit a singularity at the $T{=}U$ line. 
Here, we will complete this picture by giving more details
of the calculation of $f_{TTT}$.    
Lets us denote the function with modular weight 4 under T-duality 
 $F^{(4)}(T)$ and the function with modular weight -2 under U-duality and
the singularity at $T{=}U$ by $F^{(-2)}(U)$. Then we must have 
\begin{equation}
F^{(4)}(T) = \frac{j_T^2(T)}{j(T)(j(T)-j(i))},\;\;\;\;and\;\;\; 
F^{(-2)}(U)=\frac{j(U) (j(U) - j(i))}{j_U(U)(j(U) - j(T))}.
\label{partik101}
\end{equation}
As we can regognize $F^{(4)}(T)$ is the $E_4(T)$
function which is part of the basis of the modular forms
for the group $SL(2,Z)$. 
Moreover, $F^{(-2)}(U)$ can be rewritten in turn as 
\begin{equation}
F^{(-2)}=\frac{1}{j(T)-j(U)}\{- \frac{j_U^2(U)}{j(U)(j(U)-j(i))}\}
\frac{j_U^3(U)}{j(U)^2(j(U)-j(i))}(\frac{j_U^6}
{j^4(j(U)-j(i))^3 })^{-1}.
\label{partik102}
\end{equation}  
The terms appearing in eqn.(\ref{partik102}) represent
\begin{equation}
F^{(-2)}(U)= const.\;
\{\frac{1}{j(T)-j(U)}\}\;E_4(U)\;E_6(U) \eta^{-24}(U),
\label{partik103}
\end{equation}
where the standard theorems of modular forms predict
\begin{eqnarray}
E_4(U) \propto \frac{-j_U^2(U)}{j(j(U)-j(i))},\;E_6(U) \propto 
\frac{J_U^3(U)}{
j^2(U)(j(U)-j(i))},\;
\eta^{24} \propto \frac{j_U^6(U)}{j^4(U)(j(U)-j(i))}.
\label{partik104}
\end{eqnarray}
The function $E_4$ has a zero at $T{=}\rho$, and $E_6$ has a zero 
at $T{=}1$.
In addition, $\eta(T)$ is the well known cusp form the Dedekind
function. It has a zero at $T{=}\infty$.
Examination of the integral representation of behaviour of $f_{TTT}$ 
near the point $T{=}U$, shows that it has a single pole with 
residue $\frac{-2i}{\pi}$. This therefore 
fixes the   
numerical coefficient in front of $F^{(-2)}(T) \times F^{(4)}(U)$. 
Together with $F^{-2}(U)$, $F^{4}(T)$, 
we get the correct result eqn.(\ref{partik10}). 

{ \em The prepotential $f_{UUU}$ }

The prepotential function for the $f_{UUU}$ can be obtained by the 
replacement $T \leftrightarrow U$ but we prefer to 
find it from the equation for $f_{UUU}$. By taking appropriate
derivatives on (\ref{partik7})
we find
\begin{equation}
\partial_{U}^3f = f_{UUU}= -i(T-{\bar T})^2 \partial_{\bar T}D_U 
\partial_U {({\cal I})}_{holomorphic},
\label{oux1}
\end{equation}
or equivalently
\begin{equation}
\partial_{U}^3 f = i (T- {\bar T})^2 \partial_{T} 
D_U \partial_U {({\cal I})}_{holomorphic}.
\label{ouxc1}
\end{equation}
Here $D_U=\partial_U +\frac{2}{U-{\bar U}}$, the covariant derivative with
respect to U variable. 
It transforms with modular weight 2 under $SL(2,Z)_U$ modular transformations,
namely
\begin{eqnarray}
U\stackrel{SL(2,Z)_U}{ \rightarrow} \frac{ aU + b}{cU + d},\;\;\;\;D_U  
\rightarrow (cU+d)^2 D_U.
\label{covartibatt}
\end{eqnarray}

We should notice here, that because of the exchange symmetry $T
\leftrightarrow U$, the result for $f_{UUU}$ may come directly from
(\ref{partik10}), by the replacement $T \rightarrow U$. However, this 
can be confirmed by the solution of (\ref{oux1}). Note that the equation 
for 
$f_{UUU}$ comes from the derivative action on the holomorphic part
of the K\"ahler metric. 
By using the explicit form of the expression (\ref{metrik1}) 
and the values
of the lattice momenta (\ref{partik5}), we evaluate the 
right hand side of (\ref{oux1}) as 
\begin{equation}
f_{UUU} = 8 \pi^2 \frac{(T-{\bar T})}{(U-{\bar U})^2} \int 
\frac{d^2 \tau}{\tau_2^2} {\bar F}({\bar \tau}) \partial_{\bar \tau}
\left( \tau_2^2 \partial_{\tau} (\tau_2^2 \sum_{P_L, P_R}P_L P_R P_R P_R
e^{\pi i \tau |P_L|^2} e^{-\pi i {\bar \tau} |P_R|^2} )\right).
\label{korikou1}
\end{equation}
Further integration by parts, with the boundary term vanishing away from 
the enhanced symmetry points, gives  
\begin{equation}
f_{UUU}= 4 \pi^2 \frac{(T-{\bar T})}{(U-{\bar U})^2} \int 
\frac{d^2 \tau}{\tau_2^2} {\bar F}({\bar \tau}) 
\sum_{P_L, P_R} P_L P_R P_R P_R
e^{\pi i \tau |P_L|^2} e^{-\pi i {\bar \tau} |P_R|^2}.
\label{korikou2}
\end{equation} 
We must keep in mind that, after performing the integration in
(\ref{korikou1}, \ref{korikou2}), to keep only the
holomorphic part of the
integration and not its complex conjugate.
Using now, the modular transformations of the momenta
\begin{eqnarray}
(P_L, {\bar P}_R)\stackrel{SL(2,Z)_T}{\rightarrow} \left({\frac{cT+d}{
c{\bar T}+d}}\right)^{\frac{1}{2}} (P_L, {\bar P}_R),\;\;
(P_L, {\bar P}_R)\stackrel{SL(2,Z)_U}{\rightarrow}
\left({\frac{cU+d}{
c{\bar U}+d}}\right)^{\frac{1}{2}} (P_L, {\bar P}_R),
\label{oux5}
\end{eqnarray}
we can see that $f_{UUU}$ transforms correctly as it should, that is
modular weight 4 under $SL(2,Z)_U$ and $-2$ under $SL(2,Z)_T$. 
In addition, we can observe, in analogy with $f_{TTT}$,  
that (\ref{korikou2}) possesses a simple pole at $T=U$. 
The $f_{UUU}$ function has similar modular properties,
equivalent under the exchange monodromy transformation symmetry 
of f, and singularity structure as $f_{TTT}$.

As a result, $f_{UUU}$ takes the form
\begin{equation}
f_{UUU} = \frac{ (j(T) - j(i)) j(T)}{j_T(T) (j(U) - j(T))} \Psi(U),
\label{srty1}
\end{equation}
with $\Psi(U)$ a meromorphic modular form of weight 4 in U.
The T dependent part of $f_{UUU}$ is a meromorphic modular form of
weight $-2$ and has a singularity at the point $T=U$.

From the integral representation of $f_{UUU}$, eqn.(\ref{korikou1})
we can see that at the limit $T \rightarrow \infty$, $f_{UUU} 
\rightarrow 0$,
which means that $\Psi(U)$ is holomorphic anywhere.
Finally, we get
\begin{equation}
f_{UUU} = - \frac{2i}{\pi} \frac{ (j(T) - j(i)) j(T)}{j_T(T) (j(U) - j(T))}
\frac{j_U^2(U)}{j(U) - j(i)}\frac{1}{j(U)}.
\label{srty2}
\end{equation}
Lets us check the behaviour of (\ref{srty2}) away and at the fixed
points.
Away form the fixed points, e.g when $U \rightarrow {\hat T}= 
\frac{ a T + b}{c T + d}$, $f_{UUU}$ exhibits a singularity 
\begin{equation}
f_{UUU} \rightarrow - \frac{2i}{\pi} \frac{1}{U -{\hat T}}( cT + d)^2,
\label{srty3}
\end{equation}
such that the one loop K\"ahler metric $G_{U{\bar U}}^{(1)}$ behaves 
exactly
as expected\footnote{From the symmetry enhancement point of view for the
$SL(2,Z)$ modular group of the two torus.}, namely
\begin{equation}
G_{U {\bar U}}^{(1)} \rightarrow \frac{1}{\pi} \ln |U- {\hat T}|^2
G_{U{\bar U}}^{(0)}.
\label{srty4} 
\end{equation}
Exactly, when T is one of the fixed points of the modular group
$SL(2,Z)$, $f_{UUU}$ vanishes.
The presence of the logarithmic singulariry in the one loop corrections
to f gives rise to
the generation
of the discrete shifts in the theta angles due to monodromies around
semi-classical singularities in the quantum moduli space where
previously massive states become massless \cite{seiwi1,wkll,NS}.

{\em The one loop prepotential f}
\newline
The one-loop contribution to the K\"{a}hler metric can be
calculated through an equation different than eqn.(\ref{partik777}).
Expanding (\ref{partik7}) appropriately, we get 
\begin{equation}
2i{\cal I} =  \partial_T\partial_U f(T,U) -
\frac{2}{T- {\bar T}}f(T,U) -
\frac{2}{U - {\bar U}}f(T,U) + \frac{4}{ (T-{\bar T}) (U -{\bar U})}f(T,U)
+h.c
\label{oux11}
\end{equation}
or 
\begin{equation}
2i{\cal I} =\frac{4}{(T- {\bar T})(U - {\bar U})}(f + {\bar f})+ \dots
\label{oux1112}
\end{equation}

Acting with the appropriate derivatives on the left hand 
of (\ref{oux11}) the following identity holds:
\begin{equation}
(2f)= i(T-{\bar T})^2 (U-{\bar U})^2 
\partial_{\bar U}\partial_{\bar T}{({\cal I})}_{holomorphic},
\label{oux2}
\end{equation}
or in the symmetric form
\begin{equation}
(2f)= i(T-{\bar T})^2 (U-{\bar U})^2 
\partial_{\bar T} \partial_{\bar U}{({\cal I})}_{holomorphic}.
\label{oux222}
\end{equation}
Note that on the right hand side we kept only the holomorphic part of the 
one loop K\"ahler metric.
Explicitly, 
\begin{equation}
(2)(f + h.c) =i(T-{\bar T})^2 (U-{\bar U})^2 
\partial_{\bar U}\partial_{\bar T} \int_{\cal F} \frac{d^2\tau}{{\tau}^2_2} 
\partial_{\bar \tau}(\tau_2 Z) {\bar F}({\bar \tau}).
\label{ouxx2}
\end{equation}
Eqn's (\ref{oux2}, \ref{oux222}, \ref{ouxx2}) are equivalent and represent 
the {\em master} equation for the prepotential.
Each one of them can calculate the one loop
prepotential of any, rank four, four dimensional $N=2$ heterotic string
compactifications.
As we can observe, 
the one loop correction to the holomorphic prepotential comes by 
taking derivatives of ${\cal I}$ with respect to the conjugate 
moduli variables
from which the holomorphic prepotential does not have any dependence.
The holomorphic prepotential is defined projectively, by taking the
action
of the conjugate moduli derivatives on the holomorphic part of the one
loop
K\"ahler metric integral $\cal I$. 
In this way, we 
{\em always produce} the differential equation for the f function 
from the string amplitude. In addition, the solution of this equation 
calculates the one loop correction to the K\"ahler metric.  

The integral representation of (\ref{oux2}), after using the
explicit form of momenta (\ref{partik5}), is
\begin{equation}
f + h.c = -4 \pi (T-{\bar T}) (U-{\bar U}) \int \frac{d^2 \tau}{\tau_2^2}
{\bar F}({\bar \tau}) \partial_{{\bar \tau}} [ \tau_2^2 
\partial_{ {\bar \tau}} (\tau_2 \sum_{P_L, P_R}{\bar P}_L {\bar P}_L 
e^{i\pi\tau |P_L|^2} e^{-i\pi{\bar \tau}| P_R|^2}
)],
\label{oux331}
\end{equation}
where we have used the identity
\begin{equation}
\partial_{\bar T} \partial_{\bar U}  
Z = -\frac{4 i \pi \tau_2}{(T -{\bar T})
(U - {\bar U})} \partial_{{\bar \tau}} ( \tau_2 \sum_{P_L, P_R}{\bar
P}_L {\bar P}_L Z)
\label{oux332}
\end{equation}
and the relations
\begin{equation}
\partial_{\bar T} {\bar P}_L^2= \frac{{\bar P}_L P_R}{T -{\bar
T}}= \partial_{\bar T} {\bar P}_R^2.
\label{sxesi1}
\end{equation}
We can easily see that the one loop prepotential has the correct modular
properties, it transforms with modular weight $4$ in T and $-2$ in U.
Eqn.(\ref{oux2}) is the differential equation that the one loop
prepotential
satisfies. The solution of this equation determines 
the one loop correction
to the K\"ahler metric and the K\"ahler potential for $N=2$ orbifold 
compactifications of the heterotic string. 

Compactifications of the heterotic string on $K_3 \times T_2$,
appears to have the same moduli dependence on T and U moduli, for 
particular classes of models\cite{afgnt,kava,klm1,alta,dose1}.
Formally, the same routine procedure, namely {\em taking the 
derivatives with 
respect to the conjugate T and U moduli on} ${\cal I}$, can be applied 
to any heterotic string amplitude between two moduli scalars and  
antisymmetric tensor, in order to isolate from the general solution 
(\ref{oux2}) the term $f(T,U)$. 
The solutions for $f_{TTT}$ in 
eqn.(\ref{partik10}) and $f_{UUU}$ in eqn.(\ref{srty2}), were derived, for
$N=2$
compactification of the  
heterotic strings in \cite{afgnt} and in this work respectively, via the
modular properties of the one 
loop 
prepotential coming from the study of their integral representations.
Specific application for the model based on the 
orbifold
limit of $K_3$, namely $T^4/Z_2$, equivalent to the $SU(2)$
instanton embedding $(24,0)$, was given in \cite{mouha}. 
At the orbifold limit of $K_3$ compactification of the heterotic string 
the Narain lattice was decomposed into the form $\Gamma^{22,6}=
\Gamma^{2,2} \oplus \Gamma^{4,4} \oplus \Gamma^{16,0}$. 
It was modded by a $Z_2$ twist on the $T^4$ part together with a $Z_2$ 
shift $\delta$
on the $\Gamma^{(2,2)}$ lattice. For reasons of level matching 
$\delta^2$ was chosen to be $1/2$. The unbroken gauge group for this model
is $E_8 \times E_7 \times U(1)^2 \times U(1)^2$.
By an explicit string loop calculation via the one loop gauge couplings
the authors of \cite{mouha}, from where the one loop prepotential was 
extracted with
an ansatz, were able to calculate the third derivative of the
prepotential. The latter result was found to agree with the 
corresponding calculation in \cite{wkll,afgnt},
 which was calculated for the S-T-U subspace  
of the vector multiplets of the orbifold compactification of the 
heterotic string.
Here, we will check this result.
In particular, we will confirm the moduli dependence of the 
prepotential f on  
the trilogarithm, for the case of $SU(2)$ instanton 
embeddings \cite{lla1,fors}
$(12,12)$, $(10,14)$, $(11,13)$ and $(24,0)$ for which the 
index in the Ramond sector takes the same value,
by direct calculation of f from its master equation (\ref{oux2}).

In reality, ${\bar F}({\bar \tau})$ is the trace of ${F}^{\prime} 
(-1)^{F^{\prime}} q^{L_o - 
\textstyle{\frac{c}{24}}} {\bar q}^{{\bar L}_o -
\textstyle{\frac{c}{24}}}/{\eta}({\bar \tau})^2$ over the Ramond 
sector boundary conditions of the remaining superconformal blocks.
For the S-T-U model with instanton embedding $(d_1,d_2)=(0,24)$ 
their supersymmetric index was calculated
 in \cite{mouha}
in the form
\begin{eqnarray}
Index = \frac{1}{\eta^2} {Tr_R} F^{\prime}{(-1)}^{F^{\prime}} q^{L_o - 
\frac{c}{24}}
{\bar q}^{{\bar L}_o -\frac{c}{24}}\; =\; -2i \frac{{\bar E_4}\;
{\bar E_6}}{{\bar \Delta}},\;\;\;
\frac{{\bar E_4}{\bar  E_6}}{{\bar \Delta}}= \sum_{n \geq -1} c_1 (n) 
{\bar q}^n.
\label{partik233}
\end{eqnarray}
where $F^{\prime}$ is the right moving fermion number. 
This is exactly, the value of our index in eqn.(\ref{partiku2341}) except
for our normalization 
factor of $1/(2\pi)^2$ which accounts for the linear representation for
the dilaton.
Expanding $\cal I$ we get that
\begin{equation}
{\cal I}=(-i \pi)\int \frac{d^2 \tau}{\tau_2}(p_R^2 -\frac{1}{2\pi \tau_2})
{\bar F}({\bar \tau}).
\label{polik1}
\end{equation}
We remind here, a general remark, that the index ${\bar F}$ was  
determined using, 
the theory of modular forms, its modular properties and singularity 
structure alone.

{\em Comments on the modular integral calculation}
\newline
Let us apply eqn.(\ref{ouxx2}) for the calculation of prepotential in the 
S-T-U model. 
Remember that the prepotential for this model was calculated \cite{mouha}
from an ansatz 
solution. 
The index for this model is independent\cite{dose1} of the 
particular instanton embedding $(n_1, n_2)$ in the two $E_8$ factors
and is equal to (\ref{partik233}). We set 
\begin{equation}
\frac{E_4 E_6}{\triangle}({\bar \tau})= \sum_{n \geq -1}c(n)q^n=
c(-1)q^{-1} + c(0) + \dots
\label{tope1}
\end{equation}
The $\cal I$ integral in eqn.(\ref{oux2}) has been discussed before in
\cite{fors}. 
Using the values of the momenta (\ref{partik5})
in (\ref{oux2}) and using Poisson resummation we get
\begin{equation}
{\cal I}= (i \pi)T_2^2 \int 
\frac{d^2 \tau}{\tau_2^4} \sum_{n_1, n_2, l_1, l_2}
Q_R {\bar Q}_R e^{-2 \pi i {\bar T} det A}
e^{\frac{- \pi T_2}{\tau_2 U_2}|n_1 \tau + n_2 U \tau -U l_1 +l_2|^2}
{\bar F}({\bar \tau})
\label{tope2}
\end{equation}
where 
\begin{equation}
Q_R = \frac{1}{\sqrt{2 T_2 U_2}}(n_2 {\bar U}\tau+ n_1 \tau - {\bar U} 
l_1 + l_2),\;\;\;
{\bar Q}_R = \frac{1}{\sqrt{2 T_2 U_2}}(n_2 U \tau + n_1 \tau - U  l_1 +
l_2).
\label{tope3}
\end{equation} 
The integral (\ref{tope2}) can be calculated using the method of
decomposition into modular orbits \cite{dkl2,kikou} of $PSL(2,Z)$.
There are three contributions to the modular integral.
The zero orbit $A=0$, the degenerate orbit and the non-degenerate
orbit.
The zero orbit $A =0$ gives no contribution to the $\cal I$ integral.
The next orbit that we will examine is the
non-degenerate orbit for which the matrix representative
is
\begin{equation}
A= \pm \left( \begin{array}{cc}
k&j\\
0&p
\end{array}\right),\;\; 0 \leq j < k,\;\;p \neq 0
\label{tope4}
\end{equation}
This integral has been calculated in \cite{fors} and it is
given\footnote{After proper
incorporation of the normalization factors of our Ramond index.}
by
\begin{equation}
{\cal I}= -\frac{1}{(2 \pi)} \sum_{ \begin{array}{c}
k >0\\
l\in Z
\end{array}} \sum_{p>0} \delta_{n,kl}( \frac{2 kl}{p} +
\frac{l}{ \pi T_2 p^2} + \frac{k}{ \pi U_2 p^2} + \frac{1}{2 \pi^2 
 T_2 U_2 p^3} ) x^p + h.c,
\label{tope5}
\end{equation}
where $x \stackrel{def}{=} e^{2 \pi i (kT + lU)}$ and 
${\bar x} \stackrel{def}{=} e^{-2 \pi i(k {\bar T} + l {\bar U})}$.
Substituting eqn.(\ref{tope5}) in the {\em master equation }for the 
prepotential (\ref{oux2}) we get that the contribution of the 
orbit ${\cal I}_1$ in  f is
\begin{equation}
f|_{non-degenerate} = (2i) \left( \frac{2}{(2 \pi)^3} \sum_{(k,l)>0} c(kl)
{\cal L}i_3[e^{2 \pi i ( kT + lU)} ]\right).
\label{tope6}
\end{equation}
This is exactly the moduli dependence on the trilogarithm found 
indirectly in \cite{mouha}.
The dependence of the solution in i, out of the parenthesis in
(\ref{tope6}) is
necessary since it is used to cancel the overall dependence on
i in the one loop K\"ahler metric (\ref{partik4}).  
Note that in the previous equation we have not considered the
complex conjugate solutions which arise by taking the partial
derivatives with respect to the $\bar T$ and $\bar U$ variables in the
complex conjugate part of the solution of eqn.(\ref{tope5}).
There are two ways to see this. One is the mathematical point 
of view while the other clearly come from physical requirements.
The physical point is that the prepotential has to be a holomorphic 
function of the vector moduli variables.
On the other hand,
the integral $\cal I$, which comes
as a solution of the one loop K\"ahler metric in eqn.(\ref{partik4}),
includes the complex conjugate part of the action of the
two covariant derivatives on the prepotential f.
However, the solution for the prepotential 
as was defined here in eqn.(\ref{oux2}) 
comes from the general
solution of the K\"ahler metric which does not include the 
conjugate part of its solution.
Results of the integration coming from the degenerate orbit
and related matters will appeal elsewhere.

\section{One loop prepotential - perturbative aspects}

We will now discuss the calculation of 
perturbative corrections to the one loop prepotential using the
theory of modular forms.
We are interested on those heterotic strings which exhibit modular groups
similar to those appearing in the calculation of thresholds corrections
in non-decomposable \cite{erkl} $N=1$ symmetric orbifolds \cite{deko}. 
Let us expand at the moment the expression of eqn.(\ref{partiko224})  
around small values of the non moduli scalars $C_a$
\begin{equation}
F = -S(TU - \sum_i \phi^i \phi^i) + {h}(T,U,\phi^i),\;\;
\label{hfexp}
\end{equation}
or
\begin{eqnarray}
F = -d_{sij}T^i T^j S  +  {h}(T,U,\phi^i),\;\; d_{sij}= diag(+,-,\dots,-)
\nonumber\\
T^1=T,\;T^2=U,\;T^i = \phi^i,\;i \neq 1,2.
\label{hfexp2}
\end{eqnarray} 
The function ${h}$, the one loop correction to the 
perturbative prepotential,
enjoy a non-renormalization theorem, namely it receives perturbative
corrections only up to one loop order. Its higher loop corrections,
in terms of the $1/{(S+ \bar S)}$, vanish
due to the surviving of the discrete Peccei-Quinn 
symmetry to all orders of perturbation theory as a quantum symmetry.
In that case, under target space duality 
\begin{equation}
T \stackrel{SL(2,Z)_T}{\rightarrow} \frac{aT-ib}{icT+d},\;\; U 
\stackrel{SL(2,Z)_T}{\rightarrow} U,
\label{tathey}
\end{equation}
we get
\begin{equation}
h(T,U) \stackrel{SL(2,Z)_T}{\rightarrow} \frac{h(T,U)+ \Xi(T,U)}{(icT+d)^2}
\label{Ton1h}
\end{equation}
and a similar set of transformations under $PSL(2,Z)_{U}$. The net
result is that ${\partial}_{T}^{3} h^{(1)}(T,U)$ is a singled valued 
function of weight $-2$ under U-duality and $4$ under T-duality. 
The prepotential h modifies the K\"ahler 
potential, in the lowest order of expansion in
the matter fields, as
\begin{equation}
K = - \log[(S +{\bar S}) + V_{GS}] -
\log(T + {\bar T})^2 -\log(U + {\bar U})^2,
\label{Tontinar}
\end{equation}
where 
\begin{equation}
V_{GS} = \frac{ 2 ( h + {\bar h}) - ( T + {\bar T})(\partial_T h +
\partial_{\bar T} {\bar  h}) - ( U + {\bar U} )( \partial_U h +
\partial_{\bar U} {\bar h} ) }{(T + {\bar T}) ( U + {\bar U}) }
\label{klroupa1}
\end{equation}
is the Green-Schwarz term \cite{dere} which contains\footnote{
In the previous section, we have seen practically the direct calculation
of the $V_{GS}$ via the calculation of h. The latter appeared as f.
 However, our notation will be
different following the spirit of \cite{wkll} and f will be 
denoted by h.}  
the mixing of the dilaton with the moduli at one loop order.  
Remember, that for the conventions used at this section the
dilaton is defined as $ < S > = 4 \pi /g^2 - i \theta / 2 \pi$. 

The prepotential for $N=2$ orbifold
compactifications of the heterotic string \cite{walt,hemo,erkl}.
was calculated, from the use of its modular properties 
and singularity structure in \cite{wkll}.
Here, we adopt a similar approach
to calculate the prepotential of vector multiplets.
We discuss the calculation of 
the prepotential for the case where the moduli subspace of the 
Narain lattice associated with the T, U moduli exhibits a modular 
symmetry \cite{deko,blst1,blst2} group $\Gamma^o(3)_T \times \Gamma^o(3)_U$. 
The same modular symmetry group appears\cite{deko} in the $N=2$ sector
of the $N=1$ $(2,2)$ symmetric non-decomposable $Z_6$ orbifold defined 
on the lattice $SU(3) \times SO(8)$. In the third complex plane associated
with the square of the complex twist $(2,1,-3)/6$ the mass operator
for the untwisted subspace was given\footnote{We changed notation.
All moduli are rescaled by i.}
 to be
\begin{equation}
m^2= \sum_{m_1, m_2, n_1, n_2\;\in\;Z} \frac{1}{2T_2U_2}
|TU^{\prime}n_2 + T n_1 - U^{\prime}m_1 + 3m_2|^2,\;\;\;U^{\prime} = U+2.
\label{arga1}
\end{equation}
Let us forget the $N=1$ orbifold nature of the appearance
of this $N=2$ sector. Then its low energy supergravity theory 
is described by the underlying special geometry. 
The question now is if the prepotential, by calculating it using its modular 
properties and the singularity structure alone, as this was calculated
for decomposable\footnote{We use this term
in connection with the same type of modular symmetries
appearing in $N=1$ decomposable $(2,2)$ symmetric orbifold 
compactifications of the
heterotic string \cite{dhvw}.}
 orbifold compactifications of the heterotic 
string\cite{wkll}, has any type II dual realization.
We believe that is the case.
In the analysis of the map between type II
and heterotic dual supersymmetric string theories\cite{liya,klm1}  
it was shown that subgroups of the modular group do appear.
In particular some type II compactified on the Calabi-Yau three 
folds \cite{kava},
were shown \cite{klm1} to correspond 
in one modulus deformations of $K_3$ fibrations.
The modular symmetry groups appearing \cite{liya} are all connected
to the $\Gamma_o(N)_{+}$, the subgroup
of the $PSL(2,Z)$, the $\Gamma_o(N)$ group together
with the Atkin-lehner involutions $T\rightarrow \frac{-1}{NT}$.
We expect that the same prepotential, beyond describing the  
geometry of the $N=2$ sector of $Z_6$ in exact analogy to the  
decomposable case, may come form a compactification of the heterotic
string on the $K_3 \times T^2$. An argument that seems to give some support
to our conjecture was given in \cite{fava1}.
It was noted by Vafa and Witten 
that if we compactify a ten dimensional string theory on $T^2 \times X$,
where X any four manifold, acting with a $Z_2$ shift on the Narain lattice 
we get the modular symmetry group $\Gamma_o(2)_T \times \Gamma_o(2)_U$.
In this respect it is obvious that our calculation of the prepotential
may come from a shift
in a certain Narain lattice of $T^2$. We suspect that this is a $Z_3$
shift.
Furthermore, if we adopt N=2
conventions \cite{klemmos} in the study of dyon spectrum of $N=2$ 
supersymmetric Yang-Mills, the quantum symmetry groups $\Gamma_o(2)$,
$\Gamma^o(2)$ appear in $N_f=0, 2$ respectively with corresponding
monodromy groups $\Gamma^o(4)$,  $\Gamma_o(4)$.

From the mass operator (\ref{arga1}) we deduce that at the point $T=U$
in the moduli space of the $T^2$ torus of the untwisted plane, with
$n_1 = m_1=\pm 1$ and $n_2=m_2=0$, its $U(1) \times U(1)$ symmetry
becomes enhanced to $SU(2) \times U(1)$. Moreover, the third derivative
of the prepotential, with respect to the T variable, has to transform, in
analogy to the 
$SL(2,Z)$ case, with modular weights -2 under $\Gamma^o(3)_U$ and
4 under $\Gamma^o(3)_T$ dualities.
Using the theory of modular forms requires, for the
calculation of the vector multiplet prepotential
of the effective N=2 low energy theory of heterotic strings,
 the analog of $SL(2,Z)$ j-invariant, for $\Gamma^o (3)$,
 the Hauptmodul function.
This quantity is given by ${\omega(T)}$, where ${\omega(T)}$ is given 
explicitly by
\begin{equation}
\omega(T) = (\frac{\eta(T/3)}{\eta(T)})^{12}
\end{equation}
and represents the Hauptmodul
for $\Gamma^0(3)$,
the analogue of $j$ invariant for $SL(2,\bf Z)$.
It is obviously automorphic under $\Gamma^0(3)$ and possess\footnote{We
would lile to thank D. Zagier for pointing this to us.}
a double pole at infinity and a double zero at zero.
It is holomorphic \cite{shoe} in the upper complex plane and at
the points
zero and infinity has the expansions
\begin{equation}
{\omega(T)}= t_{\infty}^{-1} {\sum}^{\infty}_{{\lambda}=0}
a_{\lambda} t^{\lambda}_{\infty}\;, {a_{o} \neq 0},\;\;
{\omega(T)}= t_{o}^{-1} {\sum}^{\infty}_{{\lambda}=0}
b_{\lambda} t^{\lambda}_{o}\;, {b_{o}\neq 0}
\end{equation}                                                          
at ${\infty}$ and $0$ respectively with $ t = e^{- 2\pi T}$.
In full generality, the Hauptmodul
functions for the $\Gamma^0(p)$ are the functions\cite{apostol,kobl}
\begin{equation}
\Phi(\tau) = \left( \frac{\eta(\frac{\tau}{p})}{\eta(\tau)}\right)^r.
\label{haupt1}  
\end{equation}
 Here, $p= 2,3,5,7$ or 13 and $r=24/(p-1)$. For these values of p
the function in eqn.(\ref{haupt1}) remains modular invariant, i.e it
is a modular function\footnote{The hauptmodul functions for
the group $\Gamma_0(p)$ are represented by 
\begin{equation}
(\frac{\eta({\tau})}{\eta(p\tau)})^r.
\label{hauptas1}
\end{equation} }.
The function $\omega(T)$ has a single zero at zero 
and a single pole
at infinity. In addition, its first derivative
has a first order zero at zero, a pole at infinity and
a first order zero at $i{\sqrt{3}}$.
The modular form F of weight k of a given subgroup of the 
modular group $PSL(2,Z)=SL(2,Z)/{Z_2}$ is calculated 
from the formula 
\begin{equation}
\sum_{p \neq 0, \infty }{\nu_p} + \sum_{p=0,\infty} (width)
\times (order\;of\;the\;point) = \frac{\mu k}{12}.
\label{eksiso1}
\end{equation}
Here, $\nu_p$ the order of the function F, the lowest power in the  
Laurent expansion of F at p. The index $\mu$ for $\Gamma^o(3)$
is calculated from the expression\cite{shoe}
\begin{equation}
[\Gamma: \Gamma_o(N)] =N {\displaystyle\mathop{\Pi_{p/N}}}\;(1+p^{-1})
\label{extresi1}
\end{equation}
equal to four.
The width at infinity is defined as the smallest integer
such as the transformation $z \rightarrow ( z + \alpha )$ is in the group,
where $\alpha \in Z$. 
The width at zero is coming by properly transforming the width at infinity
at zero.
For $\Gamma^o(3)$ the width at infinity is 3 and the width at zero is 1.
The holomorphic prepotential can be calculated easily if we examine its
seventh derivative. The seventh derivative has modular weight
12 in T and 4 in U. In addition, it has a sixth order pole at the 
$T=U$ point whose coefficient A has to be fixed in order to produce the
logarithmic singularity of the one loop prepotential.
As it was shown \cite{afgnt,wkll} the one loop 
prepotential as T approaches $U_g = \frac{aU + b}{cU + d}$, where 
g is an $SL(2,Z)$ element\footnote{The same argument works for the 
subgroups of the modular group, but now there are 
additional restrictions on the 
parameters of the modular transformations. }
\begin{equation}
f \propto -\frac{i}{\pi} {\{(cU+d)T -(aU + b)\}^2} \ln(T - U_g).
\label{opli11}
\end{equation}
The seventh derivative of the prepotential is calculated to be
\begin{equation}
f_{TTTTTTT} = A \frac{\omega(U)_U^3  \omega(U)^5 (\omega^{\prime}(U))^3}
{(\omega(U) -\omega(\sqrt{3}))^2 (\omega(U) -\omega(T))^6} X(T),
\label{teleios1}
\end{equation}
where $X(T)$ a meromorphic modular form with modular weight 12 in T.
The complete form of the prepotential is
\begin{eqnarray}
f_{TTTTTTT} = A \left(\frac{[\omega(U)_U^3  \omega(U)^5 
(\omega^{\prime}(U))^3}{
{(\omega(U) -\omega(\sqrt{3}))^2 [(\omega(U) -\omega(T))^6}]}\right)
\left(\frac{\omega(T)_T^6}{\omega^2(T)\{(\omega(T)-
\omega (\sqrt{3}))^4 \}}\right).
\label{teleios2}
\end{eqnarray}

The two groups $\Gamma^o(3)$ and $\Gamma_o(3)$ are  conjugate to each 
other. If S is the generator 
\begin{eqnarray}
S= \left(\begin{array}{cc}
0&-1\\
1&0
\end{array}\right),\;\;we\;have\;\Gamma^o(3)=S^{-1} \Gamma_o(3) S.
\label{eftasa1}
\end{eqnarray}
So any statement about modular functions on one group is a statement 
about the
other. We have just to replace everywhere $\omega(z)$ by $\omega(3z)$ 
to go from a
modular function from the $\Gamma^o(3)$ to the $\Gamma_o(3)$.
In other words, the results for the heterotic prepotential with
modular symmetry group $\Gamma^o(3)$ may well be describe the
prepotential to the conjugate modular theory.

We have calculated the prepotential of a heterotic string
with a $\Gamma^o(3)_T \times \Gamma^o(3)_U \times Z_2^{T 
\leftrightarrow U}$ classical duality group.
The same dependence on the T, U moduli and its modular symmetry group
appears in the $\Theta^2$, $N=2$, sector 
of the $Z_6$ orbifold defined by the action of the 
complex twist $\Theta= exp[\frac{2 \pi i}{6}(2,3,-1)]$
on the six dimensional      
lattice $SU(3) \times SO(8)$, namely the $Z_6$-IIb, and on
toroidal compactifications of orbifold limits of $K_3$ in four
dimensions.

\section{Application to rank three $N=2$ heterotic string 
compactifications}

One important aspect of the expected duality is that
the vector moduli space of the heterotic string must coincide at the 
non-perturbative level with the tree level exact vector moduli space 
of the type IIA theory. 
For the type IIA superstring compactified on a Calabi-Yau space
X the internal $(2,2)$ moduli space has $N=2$ world-sheet supersymmetry for
the left and the right movers and
is described, at the large complex structure limit of X, by
the K\"ahler
moduli,  namely  $B+iJ \in H^2(X,C)$,
where $B+iJ = \sum_{i=1}^{h(1,1)} (B+iJ)_a e_a$ with $B_a$, $J_a$ real
numbers and $t_a =(B+iJ)_a$ representing the special coordinates and 
$e_a$ a basis of $H^2(X,C)$.

In this section we will derive the general form of the 
equation determining the prepotential for the rank three $N=2$ 
heterotic compactifications.
In particular we will examine a type II model admitting 
a heterotic perturbative dual realization. 
The heterotic model contains three moduli the dilaton S,  the graviphoton,
and 
the T moduli. It coincides with the 
corresponding Calabi-Yau dual model at its weak coupling limit.
In order for the heterotic prepotential to match its Calabi-Yau dual 
at its weak coupling limit a number of conditions are necessary, which
we will briefly review them here.
The existence of a type II dual to the weak coupling phase 
of the heterotic string is exactly the existence of the 
conditions \cite{lla}
\begin{equation}
D_{sss}=0,\;\;\;\;D_{ssi} =0\; for\;every\;i,\;,   
\label{koufame12}
\end{equation}
where D the Calabi-Yau divisors.
An additional condition originates 
from the higher derivative gravitational couplings
of the heterotic vector multiplets and the Weyl multiplet
of conformal $N=2$ supergravity. 
Such terms appear as well in the effective action of type II vacua 
and they have to match with heterotic one's due to duality.
In the large radius limit from the examination of the higher derivative
couplings we can infer the 
result that
\begin{equation}
D_a \cdot c_2(X) =24.  
\label{koufame15}
\end{equation}
The last condition represents \cite{lla} the mathematical
fact that the Calabi-Yau threefold X admits a fibration $\Phi$ such as
there is a map $X \rightarrow W$, with the base being $P^1$ and generic 
fiber the $K_3$ surface.
Furthermore, the area of the base of the fibration gives
the heterotic four dimensional dilaton.
In \cite{klm1} it was noticed that the nature of type II-heterotic sting
duality test has to come from the $K_3$ fiber structure over $P^1$
of the type IIA side.                                                                                                                                    
Confirmation of duality between dual pairs is then  
equivalent to the identification \cite{louisne}
\begin{equation}
\textstyle{{\cal F}_{IIA}={\cal F}_{IIA}(t^s, t^i) +
{\cal F}_{IIA}(t^i)={\cal F}_{het}^o(S, \phi^I) +
{\cal F}_{het}^{(1)}(\phi^I)}.
\label{koufame11}
\end{equation}
Here,
we have expand the prepotential of the type IIA in its large radius
limit, namely large $t_s$. In the heterotic side, we have the 
tree level classical contribution 
as a function of the dilaton S and the vector multiplet moduli 
$\phi^I$, in addition to the one loop correction as a function 
of only the $\phi^I$.
Several tests between dual models, using the indirect calculation
of the prepotential in \cite{mouha} have been performed 
in \cite{dose3,dose4,dose5}.

Dual pairs for which the prepotential in the type IIA theory is known 
can be mapped to the type IIB using mirror symmetry \cite{grple}.
In Calabi-Yau manifolds, special geometry is associated with
the description of their moduli spaces. In type IIB,
the $H^{2,1}$ cohomology describes the deformation of the complex
structure of the Calabi-Yau space $\cal M$. 
Let us consider the Calabi-Yau three fold defined as the zero 
locus of the hypersurface $P^4_{1,1,2,2,2}$ of degree eight. 
This model appears in the list of \cite{kava} as 
the A model and it is defined as $X_8(1,1,2,2,2)^{-86}_2$, where the 
subscripts and
superscripts denote the Betti numbers $b_{1,1}=2$ and $b_{1,2}=86$.
This model gives rise to 2 vector multiplets
and 86 + 1 hypermultiplets including the dilaton and its moduli space
can be studied using mirror symmetry\cite{grple,hoson3}.
The mirror manifold $X_8^{*}$ for this model is defined
by the Calabi-Yau three fold in the form $\{{\cal P}=0\}/Z_4^3$,
where the zero locus is
\begin{equation}
{\cal P}=z_1^8 +z_2^8 +z_3^4 +z_4^4 + z_5^4 -8\psi z_1 z_2  
z_3 z_4 z_5 - 2\phi z_1^4 z_2^4.
\label{koili}
\end{equation}
It depends on the deformation parameters $\phi$ and $\psi$.
The $Z_4^3$ symmetry acts on the coordinates as $(z_2, z_{2+m})
\rightarrow (-iz_2, iz_{2+m})$ for $m=1,2,3,$ respectively.
A good description of the moduli space is obtained by enlarging 
the group $\{{\cal P}=0\}/Z_4^3$ to the group $\hat G$ acting as
\begin{equation}
(z_1, z_2, z_3, z_4, z_5 ;\psi, \phi)\rightarrow 
(\omega^{{\tilde a}_1}z_1, \omega^{{\tilde a}_2}z_2,\omega^{2{\tilde
a}_3}z_3, \omega^{2{\tilde a}_4}z_4, \omega^{2{\tilde a}_5}z_5;
\omega^{-{\tilde a}}\psi,\omega^{-4{\tilde a}}\phi),
\label{koili1}
\end{equation}
 where $\textstyle{{\tilde a}=e^{2\pi i/8}}$, ${\tilde a}_i$ are
integers such as ${\tilde a}={\tilde a}_1 +{\tilde a}_2 +
2 {\tilde a}_3+ {\tilde a}_4 + {\tilde a}_5 $.
Modding the weighted projective spaces by the group $\hat G$ requires
modding out by the action $(\phi, \psi) \rightarrow (-\phi,{\tilde
\alpha} \psi)$.
The prepotential of the type IIB model defined on the mirror manifold
$X_8^{*}$ was calculated in \cite{anpa} form the study of the Yukawa 
couplings in \cite{can,hoso1,hoso2} as
\begin{equation}
{\cal F}^{II}= - 2t_1^2 t_2 - \frac{4}{3}t_1^3 + \dots + f^{NP} .
\label{peroa}
\end{equation}
From the form of the prepotential we can infer that the type II
model has a heterotic dual which corresponds to the particular
identification of $t_2$ with the heterotic dilaton
and $t_1$ with the heterotic T moduli.
As a result
\begin{equation}
f^{heterotic} = -2 ST^2 + f(T) + f^{non-pertur},
\label{peroa1}
\end{equation}
where $f(T)$ the one loop correction and $f^{non-pertur}$ the 
non-perturbative contributions.
The heterotic model is an S-T model, a two moduli example or rank three 
model, if someone takes into account the graviphoton.
The  
exact correspondence of $P^4_{1,1,2,2,2}$  with the 
three rank heterotic model  
is their connection via their classical T-duality group.
The two models at the weak coupling limit of the $t_2$ moduli
have the same classical duality group, $\Gamma_o(2)_{+}$.
Study of the discriminant of $P^4_{1,1,2,2,2}$ gives that
the conifold singularity should correspond to the perturbative $SU(2)$ 
enhanced symmetry point.
This fixes the momenta for the $\Gamma^{(2,1)}$ compactification lattice
of the heterotic model as
\begin{equation}   
p_L = \frac{i \sqrt{2} }{T -{\bar T}} \left(n_1 +n_2{\bar T}^2 +
2 m {\bar T}\right),\;\;
p_R = \frac{i \sqrt{2}}{T -{\bar T}} \left( n_1 +n_2 T{\bar T} +
m (T + {\bar T})\right).
\label{koilhju}
\end{equation}
with the enhanced symmetry point at level 2.
Let me discuss first the {\em master} equation for the general rank
three model, as well the dual heterotic of $P^4_{1,1,2,2,2}$.
The solution for the one loop correction to the K\"ahler metric 
reads \cite{anpa}
\begin{equation}
K_{T{\bar T}}=K_{T{\bar T}}^{(o)}\{1 + \frac{2i}{S- {\bar S}}{\cal I}+
\dots \}.
\label{atta}
\end{equation}
Using now the general form of solution for the K\"ahler metric
\begin{equation}
{\cal I} = \frac{i}{8} \left( \partial_T - \frac{2}{T -{\bar T}} \right)
\left( \partial_T - \frac{4}{T -{\bar T}}\right) f(T) + h.c,   
\label{koilikoili3}
\end{equation}
we can infer the {\em master} equation for the perturbative one loop
correction
to the prepotential as
\begin{equation}
2i  f(T) + h.c =  (T- {\bar T})^3 \partial_{\bar T} {\cal I}
\label{koilikoili4}
\end{equation}
or in alternative form
\begin{equation}
2 i f(T) = (T- {\bar T})^3 \partial_{\bar T} {({\cal I})}_{holomorphic}.
\label{koilikoili40}
\end{equation}

Here $K_{T{\bar T}}^{(o)}$ is the tree level metric $-2/(T-{\bar T})^2$
and ${\bar C}_l (\bar \tau)$ is the index of the
Ramond sector in the remaining superconformal blocks.
Note that eqn.(\ref{koilikoili4}) was derived from the general
solution for the one loop K\"ahler metric without any reference
to values of momenta for the $\Gamma^{(2,1)}$. This means that
this equation determines the prepotential for any rank three $N=2$
compactification of the heterotic string.
For example (\ref{koilikoili4}) determines the heterotic duals 
of the models B, C in \cite{klm1}, with associated modular groups
${\Gamma_o(3)}_{+}$, ${\Gamma_o(4)}_{+}$ 
and enhanced symmetry points at, 
the fixed points of their associated modular groups, 
Kac-Moody levels 3 and 4 respectively \cite{gome}.

Remember that the one loop K\"ahler metric \cite{anto3,anpa} for the 
heterotic model dual to the
type $P^4_{1,1,2,2,2}$ model reads
\begin{equation}
{\cal I}= \sum_{i=1}^{6}\int \frac{d^2 \tau}{\tau_2^{3/2}}
{\bar C}_l (\bar \tau) \partial_{\bar \tau}(\tau_2^{1/2}
\sum_{p_L,p_R \in \Gamma_l} e^{\pi i \tau |p_L|^2} e^{- \pi i
{\bar \tau}p_R^2}).
\label{koilikoili2}
\end{equation}
Using the above, direct substitution of the values  
of the $\Gamma^{(2,1)}$ compactification lattice momenta in (\ref{koilikoili2})
may give us the prepotential f in its integral representation.
Here the sum is over \cite{anpa} the different lattice sectors $\Gamma_l$,
 $m\;\in\;Z +\epsilon$, 
that are needed due to world-sheet modular invariance.

\section{Conclusions}

We have calculated the general equation which calculates directly
the one loop perturbative prepotential of $N=2$ heterotic string 
compactifications for any rank three or rank four parameter models
in eqn's (\ref{oux222}, \ref{ouxx2}) and (\ref{koilikoili4}, 
\ref{koilikoili40}) respectively. 
These heterotic string compactifications may or may not have a 
type II dual compactified on a Calabi-Yau.
In general, heterotic vacua with instanton embeddings numbers
$(12-n, 12 + n)$ on the $E_8 \times E_8$ gauge bundle are 
associated \cite{ftheo,lla1} to the elliptic fibrations over the 
Hirzebrush surfaces $F_n$.          
Especially, for the families of Calabi-Yau threefolds with
Hodge numbers $(3,243)$, associated with $K_3$ fibrations and elliptic 
fibrations, when
n is even the rank four Calabi-Yau is an elliptic fibration over   
the Hirzebrush surface $F_2$ or $F_0$, while for n odd the rank four
models are given in terms of the Hirzebrush surface $F_1$.
At the heterotic perturbative level all the models, which are coming
from complete Higgising of the charged hypermultiplets,
with the same Hodge numbers, come from the instanton embeddings
$(12,12)$, $(11,13)$, $(10,14)$. 
However, at the heterotic perturbative level all the models are the 
same as
we have already said. This is clearly seen from the nature of the
Ramond index (\ref{partiku2341}) which is independent from the
particular instanton embedding.
In particular, we tested the moduli dependence of the
prepotential, coming from the non-degenerate orbit, for the
previous $SU(2)$ instanton embeddings, and the $(24,0)$ one, against the 
moduli dependence of the prepotential extracted from the one loop
corrections to the gauge couplings in \cite{mouha}.
In addition, we calculated the differential equation of the 
third derivative, of the prepotential for the rank four S-T-U model
with respect of the complex structure U variable, and exhibit its
solution.

The {\em master} equation's (\ref{oux222}), (\ref{koilikoili40})
 open the way for direct testing 
of the web of dualities, e.g 
the duality between type I and $K_3 \times T^2$ at their weakly coupled
region which was tested via the third derivatives of the one loop
prepotential in \cite{anbac}.
However, there are other dualities which can be tested at the quantum level. 
For example, if we continue further compactification on $S^1$ of
F-theory
defined on the elliptic Calabi-Yau 3-fold, we get duality between 
M-theory \cite{var} on the associated Calabi-Yau three fold and heterotic
strings
on $K_3 \times S^1$. Further compactifying on $S^1$, we get duality between 
type IIA on 
Calabi-Yau three folds and heterotic on $K_3 \times T^2$. 
Furthermore, the 
direct way of calculating the holomorphic prepotential in  
(\ref{oux2}), (\ref{koilikoili4}) 
can calculate the $N=2$
 central charge and $N=2$ BPS spectrum as well the black hole
entropy \cite{lubl1,reybl2}.

{\bf Acknowledgments}

We are grateful to I. Antoniadis, C. Kounnas, K. S. Narain, S.
Stieberger, D. Zagier and R. Lewes for 
useful conversations and D. Bailin for reading the manuscript.

{\bf Appendix A}\setcounter{equation}{0}
\def\theequation{A.\arabic{equation}} 

{\cal A1} {\em Useful relations with modular forms}

The functions $E_4$, $E_6$ form the basis of modular forms for
the group $SL(2,Z)$ and are defined in term of Eisenstein series 
of weight four and six. Namely,
\begin{equation}
E_{2k}(T)= \sum_{
n_1, n_2 \in Z}^{\prime}
(in_1 T + n_2)^{-2k},\;\;k\;\in\;Z.
\label{partik1041}
\end{equation}
Here the prime means that $n_1 \neq 0$ if $n_2 = 0$.
Let us provide some useful relations between the basis for
modular forms for $PSL(2,Z)$ and $\triangle$
and the j invariant.
With the use of these relations various results appeared in the 
literature, like those in \cite{wkll}, \cite{afgnt} can be easily
translated to each other.
It can be proved, using the singularity stucture of the modular forms, 
that the following relations hold
\begin{equation}
E_4(T) = -\frac{(j^{\prime})^2}{4 \pi^2 j(j-j(i))} = 1+240 
\sum_{n=1}^{\infty} \sigma_3(n) e^{2 \pi i T},
\label{areko1}
\end{equation}
\begin{equation}
E_6(T) =\frac{(j^{\prime})^3}{(2 \pi i)^3 j^2(j-j(i))}=
1-504\sigma_5(n) q^n,
\label{areko2}
\end{equation}
\begin{equation}
\triangle(T)= -\frac{1728(j^{\prime})^6}{(48 \pi^2)^3 j^4 (j-j(i))^3}= 
\eta^{24}(T)= \frac{1}{(2 \pi i)^6}\frac{(j^{\prime})^6}{j^4 (j-j(i))^3},
\label{areko3}
\end{equation}
where $\eta(T)$ is the Dekekind function
and the value of $\sigma$  represents the sum over divisors
\begin{equation}
\sigma_{h-1}(n)\stackrel{def}{=} \sum_{d/n} d^{h-1}.
\label{areko4}
\end{equation}
We have used the notation
\begin{equation}
j^{\prime}(T) = j_{T}(T).
\label{areko24}
\end{equation}
Note that in general $E_4$, $E_6$ and $\triangle$ are defined 
in terms of the Klein's absolute invariant J as
\begin{equation}
j(T)\stackrel{def}{=} 1728 J(T),
\label{areko44}  
\end{equation}
\begin{equation}
J(T)= \frac{E_4^3(T)}{1728
\triangle(T)}=
1 + \frac{E_6^2(T)}{1728 \triangle(T)},\;\;\;T \in H
\label{areko5}  
\end{equation}
and 
\begin{equation}
j(T)=e^{-2 \pi i T} + 744 + 196884
e^{2 \pi i T} + \dots
\label{areko6}
\end{equation}
Remember that the following relations are valid
\begin{equation}
j(T)= \frac{E_4^3(T)}{\triangle(T)},
\label{areko7}
\end{equation}
and
\begin{equation}
j(T)= \frac{E_6^2(T)}{\triangle(T)}.
\label{areko8}
\end{equation}
Here, j is the modular invariant function for the
inhomogeneous  modular group
$PSL(2,Z)$.


\newpage
\begin{thebibliography}{700}
\bibitem{seiwi1}N. Seiberg and E. Witten, Nucl. Phys. B426 (1994) 19,   
Erratum B430 (1994) 485;\\ B431 (1994) 484.
\bibitem{club}A. Klemm, W. Lerche, S. Theisen and S. Yankielowicz;
Phys. Lett. B344 (1995)\\ 169, P. Argyres and A. Faraggi,
Phys. Rev. Lett. 73 (1995) 3931.
\bibitem{dis}R. Dijkgraaf, E. Verlinde and H. Verlinde, Com. Math. Phys. 
115 (1988) 669;\\ {\em On Moduli Spaces of Conformal Field Theories
with $c \geq 1$}, Proceedings\\ Copenhagen Conference,{\em Perspectives in
String Theory}, ed. by P. Di Vecchia and\\
 J. L. Petersen, World Scientific, 
Singapore, 1988. 
\bibitem{can}P. Candelas, X. de la Ossa, P. Green and L. Parkes,
Phys. Lett.  B258 (1991) 118;\\ Nucl. Phys. B359 (1991) 21.
\bibitem{anto3}I. Antoniadis, E. Gava and K. S. Narain and 
T. R. Taylor, Nucl. Phys. B407 (1993) 706.
\bibitem{deko}P. Mayr and S. Stieberger, Nucl. Phys.  B407 (1993) 725;D.
Bailin,
 A. Love,\\ W. A. Sabra and S. Thomas, 
Mod. Phys. Letters.  A9 (1994) 67; A10 (1995) 337;\\
C. Kokorelis, String Loop Threshold Corrections from Generalized
Coxeter Orbifolds, SUSX-TH-98-010, to appear.
\bibitem{erkl}J. Erler and A. Klemm, Commun. Math. Phys.  153 (1993) 579. 
\bibitem{fors}K. Foerger and S. Stieberger, Nucl.Phys. B514 (1998) 135.
\bibitem{dkl2}L. Dixon, V. Kaplunovsky and J.Louis,
Nucl. Phys. 355 (1991) 649.
\bibitem{dhvw}L. Dixon, J. Harvey, C. Vafa and E. Witten,
Nucl. Phys. B261 (1985) 678;\\ Nucl. Phys. B274 (1986) 285. 
\bibitem{imnq}L. E. Ib\'a\~nez, J. Mas, H. P. Nilles and F. Quevedo,
 Nucl. Phys.  B301 (1988) 157.
\bibitem{osa} P. Candelas, X. de la Ossa, A. Font, S. Katz and D. R.
Morrison,\\ Nucl. Phys.  B416 (1994) 481.
\bibitem{apostol}T. M. Apostol, Modular functions and Dirichlet series,
 (Springer 1976).
\bibitem{kobl}N. Koblitz, Introduction to Elliptic curves and modular
forms, (Springer 1983).
\bibitem{blst1}D. Bailin, A. Love, W. Sabra and S. Thomas;
Phys. Lett.  B320 (1994) 21.
\bibitem{blst2}D. Bailin, A. Love, W. Sabra and S. Thomas,
Mod.Phys. Lett.  A9 (1994) 1229.
\bibitem{kikou} E. Kiritsis and C. Kounnas, Nucl. Phys.  B442 (1995) 472.
\bibitem{mouha}J. Harvey and G. Moore,  Nucl. Phys.  B463 (1996) 315.
\bibitem{louisne}J. Louis and K. Foerger, Holomorphic couplings
in string theory, Nucl. Phys. Proc.\\ Suppl. 55B (1997) 33.
\bibitem{sv}M. A Shifman and A. I. Vainstein, Phys. Lett B 359 (1991)
571;
Nucl. Phys  B 277 (1986) 456;V. Kaplunovsky and J. Louis,  Nucl. Phys  B
444 (1995) 501.
\bibitem{kava}S. Kahru and C. Vafa, Nucl. Phys  B450 (1995) 69. 
\bibitem{fhsv}S. Ferrara, J. A. Harvey, A. Strominger, C.Vafa,
 Phys. Lett.  B361 (1995) 59.
\bibitem{ftheo}C. Vafa, Nucl. Phys.   B469 (1996) 403.
\bibitem{aspo1}P. Aspinwall, Enhanced gauge symmetries and $ K_3$
 surfaces, Phys. Lett.  B357\\ (1995) 329.
\bibitem{var}E. Witten,  Nucl. Phys.  B443 (1995) 85.
\bibitem{hastro}J. Harvey and A. Strominger, Nucl. Phys.  B449
(1995) 535.
\bibitem{wkll}B. de  Wit, V. Kaplunovsky, J. Louis and D. Lust,
Nucl. Phys.  B451 (1995) 53.
\bibitem{afgnt}I. Antoniadis, S. Ferrara, E. Gava, S. Narain,
T. R. Taylor, Nucl. Phys.  B447\\ (1995) 35.
\bibitem{walt}M. Walton, Phys. Rev.  D37 (1987) 377.
\bibitem{cgh}P. Candelas, P. Green and T. Hubsch, Phys. Rev. Lett.  62
(1989) 1956.
\bibitem{alta}G. Altazabal, A. Font, L. E. Ibanez, and F. Quevedo,
Nucl. Phys. B 461 (1996) 85
\bibitem{klm1}A. Klemm, W. Lerche and P. Mayr, Phys. Lett. B357 (1995) 
313.
\bibitem{hoso1}S. Hosono, A. Klemm, S. Theisen and S.-T.Yau, Nucl.
Phys.  B433 (1995) 501.
\bibitem{hoso2} S. Hosono, A. Klemm, S. Theisen and S.-T.Yau, Com.
Math. Phys. 167 (1995) 301. 
\bibitem{hoson3}P. Candelas, X. de la Ossa, A. Font, S. Katz and D.R.
 Morrison, Nucl. Phys.  B416, (1994) 481.  
\bibitem{fava1}C. Vafa and E. Witten, Nucl. Phys. B (Proc. Suppl.)
 46 (1996) 225.
\bibitem{lla}P. Aspinwall and J. Louis, Phys. Lett.   B369 (1996) 233;
P. Aspinwall, Phys. Lett. \\ B371 (1996) 231, Enhanced gauge
symmetries and Calabi-Yau three folds.
\bibitem{lla1}D. Morrison and C. Vafa, Nucl.Phys. B473 (1996) 74;
Nucl.Phys. B476 (1996) 437.
\bibitem{grple}B. Greene and R. Plesser, Nucl. Phys. B338 (1990) 15.
\bibitem{erle}J. Erler, J. Math. Phys. 35 (1994) 1819.
\bibitem{klemmos}A. Klemm, "On the Geometry behind $N=2$ Supersymmetric
Effective actions\\ in four dimensions", hep-th/9705131.
\bibitem{iss1}E. Witten, Nucl. Phys.  B286 (1986) 79.
\bibitem{dose1}G. L. Cardoso, G. Curio and D. Lust, Nucl.Phys.  B491
(1997) 147.
\bibitem{anpatay}I. Antoniadis, H. Partouche and T. R. Taylor, "Lectures
on Heterotic-Type I\\
 duality", hep-th/9706211.
\bibitem{dose2}V. Kaplunovshy, J. Louis and S. Theisen, Phys. Lett. 
 B357 (1995) 71.
\bibitem{dose3}G. L. Cardoso, G. Curio, D. Lust,
T. Mohaupt, Instanton Numbers and Exchange Symmetries in $N=2$ Dual
String Pairs, Phys. Lett. B382 (1996) 241.
 \bibitem{dose4}T. Kawai, String duality and Modular Forms,
Phys.Lett.  B397 (1997) 51.
\bibitem{dose5}G. L. Cardoso, G. Curio, D. Lust,
Perturbative Couplings and Modular Forms in N=2 String Models with a
Wilson Line,Nucl.Phys. B491 (1997) 147.
\bibitem{liya}B. H. Lian and S. T. Yau, Com. Math. Physics. 176 (1996) 163. 
\bibitem{hemo}M. Henningson and G. Moore, Nucl.Phys.  B482 (1996) 187.
\bibitem{anpa}I. Antoniadis and H. Partouche, Nucl.Phys.  B460 (1996)
470.
\bibitem{cefergi1}S. Cecotti, S. Ferrara and M. Villasante, Int. J. Mod.
Phys. A2 (1987) 1839.
\bibitem{kuue}T. Kugo and S. Uehara,  Nucl. Phys. B222 (1983) 125.
\bibitem{NS}N. Seiberg, Phys. Lett. B206 (1988)75.
\bibitem{dere}J. P. Derendinger, S. Ferrara, C. Kounnas and F. Zwigner, 
Nucl. Phys.  B372 (1992)\\ 145.
\bibitem{sukou}E. Cremmer, C. Kounnas, A. Van Proyen, J. P. Derendinger,
S. Ferrara,\\
 B. de Wit and L. Girardello, Nucl. Phys.  B250 (1985) 385.
\bibitem{shoe}B. Shoeneberg, Elliptic modular functions, (Hamburg, 1974). 
\bibitem{DWLVP} B. de Wit, P. Lauwers and A. Van Proeyen,
  Nucl. Phys.  B255 (1985) 569.
\bibitem{gome}C. Gomez, R. Hernandez and E. Lopez, Nucl. Phys. B501
(1997) 109.
\bibitem{anbac}I. Antoniadis, C. Bachas, C. Fabre, H.
Partouche and
 T. R. Taylor,\\ Nucl. Phys. B498 (1997) 160.
\bibitem{lubl1}K. Behrndt, G. Lopes Cardoso, B. de Wit, R. Kallosh, 
D. Lust and\\ T. Mohaupt, Nucl.Phys.  B488 (1997) 236.
\bibitem{reybl2}S. J. Rey, Nucl. Phys.  B508 (1997) 569.
\end{thebibliography}
\end{document}